\documentclass[pre,amssymb,amsmath,preprintnumbers,nofootinbib, 11pt]{revtex4-1}
	
	\pdfoutput=1
	\usepackage{graphicx}
	\usepackage{color}
	\usepackage{verbatim}
	\usepackage{subfigure}
	\usepackage{amssymb}

\begin{document}

	\title{A proof by graphical construction of the no-pumping theorem of stochastic pumps}
	\author{Dibyendu Mandal$^1$ and Christopher Jarzynski$^2$}
	\affiliation{
	$^1$Department of Physics, University of Maryland, College Park, MD 20742, U.S.A.\\
	$^2$Department of Chemistry and Biochemistry, and Institute for Physical Science and Technology,University of Maryland, College Park, MD 20742, U.S.A.
	}

	\begin{abstract}
	A {\it stochastic pump} is a Markov model of a mesoscopic system evolving under the control of externally varied parameters.
	In the model, the system makes random transitions among a network of states.
	For such models, a ``no-pumping theorem'' has been obtained, which identifies minimal conditions for generating directed motion or currents.
	We provide a derivation of this result using a simple graphical construction on the network of states.
	\end{abstract}

	\maketitle

	\section{Introduction}
	\label{sec:intro}
	
	The term ``molecular motors'' refers to subcellular molecular complexes that perform biologically important tasks such as carrying loads in intracellular  transport, contracting muscle cells and polymerizing microtubules \cite{Howard2001,Sperry2007}.  Because of their small size, molecular motors are strongly influenced by the thermal fluctuations of the surrounding medium, and thus exhibit highly stochastic behavior.  In recent years there has been considerable and growing interest in synthesizing artificial analogues of these molecular motors.  Achievements to date include molecular walkers on a DNA origami track \cite{Pei2006, Lund2010, Simmel2009}, rotating catenanes \cite{Leigh2003}, nanoscale assembly lines \cite{Gu2010}, and single-molecule electric motors \cite{Tierney2011}.
	
	Unlike their biological counterparts, artificial molecular machines are generally non-autonomous: they are manipulated by varying external parameters or stimuli such as temperature, chemical environment, or laser light.  In view of this it becomes interesting to investigate, from a general theoretical perspective, how a small system evolving in a thermal environment can be controlled by means of externally driven parameters.  In this context the term ``stochastic pump'' has come to denote a model class of systems (specified more precisely in Section~\ref{sec:setup} below) whose dynamics are characterized by random transitions among a discrete set of states, as the transition rates themselves are varied externally.  Stochastic pumps capture essential features of non-autonomous molecular machines while remaining amenable to exact mathematical analysis.
	
	Experiments by Leigh and coworkers \cite{Leigh2003} on {\it catenanes} -- mechanically interlocked, ring-like molecules -- provide a paradigm of non-autonomous, artificial molecular machines that can be modeled as stochastic pumps~\cite{Astumian2007,Rahav2008,Chernyak2009}.  In these experiments, one or two small rings make transitions among a set of binding sites on a large ring.  These transitions can be treated as a Poisson process, as in Sec. \ref{sec:setup} below.  By using laser light and changes in temperature to perturb the conformations of these binding sites, the transition rates can be manipulated in a time-dependent manner.
	
	The study of stochastic pumps focuses on the flow of probability that arises in response to the time-dependent pumping of the external parameters.  In the specific example of the catenane experiments of Ref.\ \cite{Leigh2003}, this flow of probability describes the statistics of the motion of the small ring (or rings) from one binding site to another on the large ring.  For the case of adiabatic (quasi-static) pumping, the generation of such probability currents can be understood in terms of geometric phases~\cite{Astumian2007,Sinitsyn2007b,Chernyak2009,Sinitsyn2009}, analogous to Berry's phase in quantum mechanics~\cite{Berry1984}.  For the more general case of non-adiabatic stochastic pumps, a ``no-pumping theorem'' has been obtained, specifying conditions under which the time-periodic driving of a stochastic pump leads to no net flow of probability~\cite{Rahav2008, Chernyak2008, Horowitz2009, Maes2010}.  This result will be the main focus of this paper.
		
	Rahav {\it et al}~\cite{Rahav2008} derived the no-pumping theorem by analyzing properties of matrices, and Chernyak and Sinitsyn~\cite{Chernyak2008} showed that this result follows from a quite general ``pumping restriction theorem'' related to the topology of the stochastic pump.
	Horowitz and Jarzynski~\cite{Horowitz2009} extended the result to one-dimensional Brownian models.
	More recently, Maes {\it et al}~\cite{Maes2010} have obtained and extended the no-pumping theorem by considering the embedded Markov chain associated with the stochastic pump.
	The aim of the present paper is to provide a quick derivation of the no-pumping theorem using an elementary graph theoretic construction. We first give an introduction to our mathematical set up and a brief statement of the no-pumping theorem (Sec. II), then we illustrate our derivation with a simple example (Sec. III) and finally we give the general proof (Sec. IV).

	\section{setup}
	\label{sec:setup}
	
\begin{figure}[tbp]
	\includegraphics[scale = .75, angle = 0]{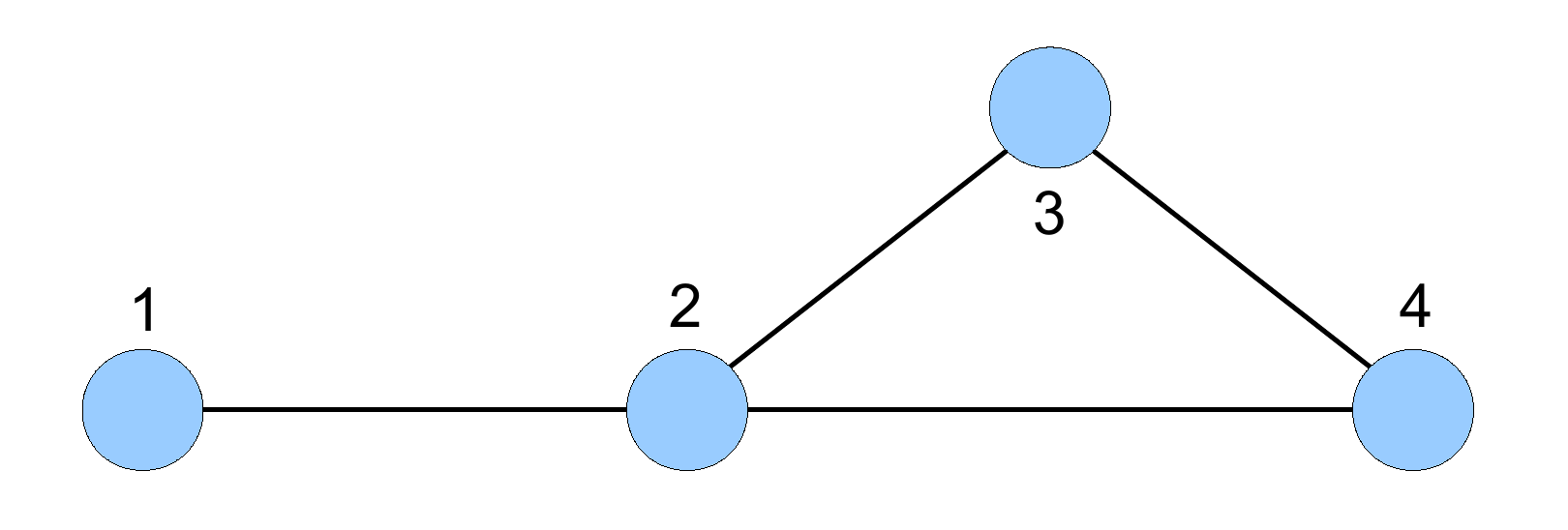}
	\caption{Graphical representation of a 4 state system with a single cycle $\{2,3,4\}$.}
	\label{fig:example}
	\end{figure}

	Consider a system whose evolution is modeled by random jumps among a set of $N$ states.  Specifically, we will model these jumps as a Poisson process: if the system is in state $j \in \{ 1, \ldots, N\}$, its rate of jump to some other state $i$ is given by a real number $R_{ij} \geq 0$.  We assume  $R_{ji} \neq 0$ if $R_{ij} \neq 0$, but the two rates need not be equal. It is convenient to represent these states and jumps by a graph $G$ with $N$ vertices and $E$ edges.  Each vertex represents one state, and each edge indicates positive transition rates between the pair of vertices it connects ($R_{ij}, \, R_{ji} > 0$).  Thus the dynamics are fully specified by $2E$ positive transition rates $R_{ij}$ ($i\ne j$).  This is illustrated in Fig. \ref{fig:example} for a system with $N = 4$ states and $E = 4$ edges. We assume $G$ to be connected, in the sense that any vertex can be reached from any other by following some sequence of edges. Let us also define a {\it cycle} of a graph to be an ordered set of more than two vertices, with edges between consecutive elements, and between the first and last elements.  Thus a cycle can be pictured as a closed loop formed by a sequence of edges in the graph. There is one such cycle $\{2,3,4\}$ in Fig. \ref{fig:example}. 

	Let $p_i(t)$ denote the probability to find the system in state $i$ at time $t$. The instantaneous probability current from state $j$ to state $i$, denoted by $J_{ij}(t)$, is then given by
	\begin{equation}
	\label{eq:current}
	J_{ij}(t) = R_{ij} p_j(t) - R_{ji} p_i(t)
	\end{equation}
	which is anti-symmetric with respect to exchange of indices: 
	\begin{equation}
	\label{eq:antisymmetry}
	J_{ji} = - J_{ij}.
	\end{equation}
	The rate of change of the probability to find the system in any state $i$ is the net current into that state:
	\begin{equation}
	\label{eq:evolution}
	\dot{p_i}(t) = \sum_{j \neq i} J_{ij}(t).
	\end{equation}
	For the system in Fig. \ref{fig:example} these are explicitly written down in equation \eqref{eq:evolution_example}.
	
	For a system that satisfies the above assumptions let us first consider the case of fixed transition rates $R_{ij}$.  The state probabilities are then described by a vector ${\bf p}(t) = (p_1,\cdots,p_N)$ that evolves asymptotically toward a unique steady-state distribution ${\bf p}^s$~\cite{Schnakenberg1976,vanKampen2007}.  The transition rates $\{ R_{ij} \}$ are said to satisfy detailed balance if $J_{ij}^s \equiv R_{ij}p_j^s - R_{ji}p_i^s = 0$ for every pair $(i,j)$, that is, if all probability currents vanish in the steady state.  In this scenario, the non-zero transition rates can be written in the form
	\begin{equation}
	\label{eq:Arrhenius}
	R_{ij} = e^{-(B_{ij}-E_j)}
	\end{equation} 
	with (crucially)
	\begin{equation}
	\label{eq:symmetry}
	B_{ij} = B_{ji}
	\end{equation}
	for all $(i,j)$.
	Specifically, if we let $E_i \equiv -\ln p_i^s$ denote the ``stochastic potential'' of state $i$ \cite{Ge2010, Kubo1973}, then the condition for detailed balance becomes $R_{ij}e^{-E_j} = R_{ji}e^{-E_i}$.  Comparing with equation \eqref{eq:Arrhenius}, we see that this condition is equivalent to the symmetry $B_{ij} = B_{ji}$ (equation \eqref{eq:symmetry}).  Because of the evident similarity between equation \eqref{eq:Arrhenius} and the familiar Arrhenius expression for thermally activated transitions~\cite{Astumian2007}, we will interpret the $E_i$'s as the effective free energies of the $N$ states and the $B_{ij}$'s as effective free energy barriers between them.

	Let us now move on to the case of time-dependent transition rates $R_{ij}(t)$, and let us henceforth assume that these rates satisfy detailed balance at all times.  (In other words, if we were to ``freeze'' the rates at their values at any instant in time, then the system would subsequently relax to a steady state with zero currents, $J_{ij}^s = 0$ for all $i$, $j$.)  We can then interpret these time-dependent rates as arising due to state and barrier energies that vary with time, $\{ E_i(t) \}$ and $\{ B_{ij}(t) \}$.  If these variations are periodic in time, say with period $T$, then according to Floquet theory \cite{Talkner1999}, the system response also becomes periodic  in the limit of long time: $p_i(t+T) = p_i(t)$. We will call this the {\it periodic steady state} and will denote the corresponding quantities by a superscript $ps$.
	
	We denote by $\Phi_{ij}^{ps}$ the integrated probability current from state $j$ to state $i$ over a time period $T$ in periodic steady state, i.e. 
	\begin{equation}
	\label{eq:integrated_current}
	\Phi_{ij}^{ps} = \int_T \mathrm{d}t \, J_{ij}^{ps}(t).
	\end{equation}
	From equation \eqref{eq:antisymmetry} these are also antisymmetric:
	\begin{equation}
	\label{eq:antisymmetryPhi}
	\Phi_{ji}^{ps} = - \Phi_{ij}^{ps}.
	\end{equation}
	These integrated currents are the objects of our interest, as they reveal whether or not the periodic pumping of the state and barrier energies produces a {\it directed} flow of probability throughout the network of states.
	If $\Phi_{ij}^{ps} \ne 0$ for some $(i,j)$, then this indicates a net flow of probability, over each period of pumping, along the edge connecting states $i$ and $j$.
	Conversely, if $\Phi_{ij}^{ps} = 0$ for every edge in the graph, then the probability currents $J_{ij}^{ps}(t)$ might slosh back and forth, so to speak, but there is no net circulation of current.
	
	In the context of the catenane experiments mentioned earlier \cite{Leigh2003}, the directed flow of probability is manifested in the unidirectional rotation of the small ring(s) around the large ring.  Indeed, the experimentally observed absence of unidirectional rotation in the case of [2]-catenanes (one small ring interlocked with one large ring) is an instance of the no-pumping theorem, which we now state explicitly.
	
	With the above definitions and assumptions in place, the no-pumping theorem asserts that \emph{if} either all the state energies $\{E_i\}$ or all the barrier energies $\{ B_{ij}\}$ are kept fixed in time during the pumping, \emph{then} the integrated probability current is zero along all edges, i.e. 
	\begin{equation}
	\label{eq:NPT}
	\Phi_{ij}^{ps} = 0 \hspace{.1 in} \text{for all pairs} \hspace{.1 in} (i,j).
	\end{equation}
	Consequently one must vary at least one state energy \emph{and} at least one barrier energy to produce directed probability currents in the periodic steady state.
	
	The case of fixed state energies $\{ E_i \}$ and time-dependent barriers $\{ B_{ij}(t) \}$ is straightforward: the system asymptotically approaches a fixed steady-state distribution $p_i^s = \exp{(-E_i)}$ \cite{Rahav2008}. Equations \eqref{eq:current}, \eqref{eq:Arrhenius} and \eqref{eq:symmetry} then imply
	\begin{eqnarray}
	\label{eq:NPT_example}
	J_{ij}^s(t) & = & e^{-[B_{ij}(t) - E_j]} e^{-E_j} - e^{-[B_{ji}(t) - E_i]} e^{-E_i} \nonumber \\
	& = & e^{-B_{ij}(t)} - e^{-B_{ji}(t)} = 0 \nonumber
	\end{eqnarray}
	for all $(i,j)$. Thus the instantaneous currents vanish, and therefore so do the integrated currents.  Hence in the following sections we focus on the less obvious case of fixed barrier energies $\{ B_{ij} \}$, but periodically pumped state energies, $\{ E_i(t) \}$.

	\section{illustration of Proof}
	\label{sec:example}
	
	Consider the system in Fig. \ref{fig:example}, with $N = 4$ states, $E = 4$ edges and a single cycle, and assume that all the $B_{ij}$'s are fixed in time while one or more of the $E_i(t)$'s are varied periodically.

	Combining equation \eqref{eq:evolution} with the antisymmetry of $J_{ij}$'s, equation \eqref{eq:antisymmetry}, we have
	\begin{equation}
	\label{eq:evolution_example}
	\begin{split}
	\dot{p_1} & =  J_{12}(t)   \\
	\dot{p_2} & =  - J_{12}(t) + J_{23}(t) - J_{42}(t)  \\
	\dot{p_3} & =  - J_{23}(t) + J_{34}(t) \\
	\dot{p_4} & =  J_{42}(t) - J_{34}(t) .
	\end{split}
	\end{equation}
	In the periodic steady state there is no net change in state probabilities over a time period $T$, i.e. $\int_T \dot{p_i}(t) \, \mathrm{d}t = 0$ for all $i$, hence
	\begin{equation}
	\label{eq:kcl_example}
	\begin{split}
	0 & =  \Phi_{12}^{ps}  \\
	0 & =  - \Phi_{12}^{ps} +  \Phi_{23}^{ps} - \Phi_{42}^{ps} \\
	0 & =  - \Phi_{23}^{ps} + \Phi_{34}^{ps}  \\
	0 & =  \Phi_{42}^{ps} - \Phi_{34}^{ps}
	\end{split}
	\end{equation}
	where we have integrated equation \eqref{eq:evolution_example} over one period of the periodic steady state. Since normalization implies $\sum_i{\dot{p_i}} = 0$, only 3 of the 4 equations in either \eqref{eq:evolution_example} of \eqref{eq:kcl_example} are independent. The solution of equation \eqref{eq:kcl_example} therefore contains a single free parameter:
	\begin{equation}
	\label{eq:kclsol_example}
	\Phi_{12}^{ps} = 0 \quad , \quad \Phi_{23}^{ps} = \Phi_{34}^{ps} = \Phi_{42}^{ps} = \Phi.
	\end{equation} 
	These results are easy to understand: $\Phi_{12}^{ps}=0$ because the edge $(1,2)$ does not belong to any cycle, and the currents along the remaining edges are equal because they all belong to the same cycle, which is the only  cycle in the graph. This intuition has been formalized and generalized to arbitrary graphs by Chernyak {\it et al} \cite{Chernyak2008}.
	
	Detailed balance implies further constraints. From equations \eqref{eq:current} and \eqref{eq:Arrhenius} we have
	\begin{equation*}
	J_{ij}(t) \, e^{B_{ij}} = e^{E_j(t)} p_j(t) -  e^{E_i(t)} p_i(t).
	\end{equation*}
	Summing both sides of this equation over the edges along the cycle $\{2,3,4\}$ then gives
	\begin{equation}
	\label{eq:db_example}
	J_{23}(t) \, e^{B_{23}} + J_{34}(t) \, e^{B_{34}} + J_{42}(t) \, e^{B_{42}} = 0.
	\end{equation}
	We have deliberately omitted the superscript $ps$ to indicate that the above relation holds whether or not the system has reached the periodic steady state. Indeed, equation \eqref{eq:db_example} remains true even if the external driving is not periodic, and even if the barriers are time-dependent. We note that a generalized form of this equation for arbitrary graphs was used by Chernyak and Sinitsyn to derive a ``pumping-quantization theorem" for integrated probability currents: in the low-temperature, adiabatic limit, each integrated current is expressed in terms of a vector potential in the space of externally controlled parameters, and exhibits quantized behavior \cite{Chernyak2009}.
		
	Returning to the periodic steady state with fixed barriers $\{ B_{ij} \}$, we integrate equation \eqref{eq:db_example} over one period $T$ to get
	\begin{equation}
	\label{eq:kvl_example}
	\Phi_{23}^{ps} \, e^{B_{23}} + \Phi_{34}^{ps} \, e^{B_{34}} + \Phi_{42}^{ps} \, e^{B_{42}} = 0.
	\end{equation}
	Combined with equation \eqref{eq:kclsol_example} this gives
	\begin{equation*}
	\Phi \, \big( e^{B_{23}} + e^{B_{34}} + e^{B_{24}} \big) = 0.
	\end{equation*}
	Hence $\Phi = 0$, and all the integrated probability currents $\Phi_{ij}^{ps}$'s in the system are zero.

	\section{General Proof}
	\label{sec:general}
	
Consider a connected graph $G$ with $N$ vertices and $E$ edges.
As before, we assume that the $2 E$ transition rates satisfy detailed balance at all times, hence they can be written in the form $R_{ij} = e^{[-(B_{ij}-E_j)]}$ with $B_{ij} = B_{ji}$. We now imagine that the state energies $E_i(t)$ are varied periodically with time, while the barriers energies $B_{ij}$ are held fixed. After the system has reached a periodic steady state, ${\bf p}(t+T) = {\bf p}(t)$, integration of equation \eqref{eq:evolution} over one period yields
	\begin{equation}
	\label{eq:kcl}
	\sum_{j \neq i} \Phi_{ij}^{ps} = 0 \quad \text{for all} \, i.
	\end{equation}
	As with equation \eqref{eq:kcl_example} only $(N-1)$ of these $N$ equations are independent. Moreover, equation \eqref{eq:kcl} implies that if $\Phi_{ij}^{ps} > 0$ for a connected pair of states $(i,j)$, then there must exist at least one other vertex $k$ such that $\Phi_{ik}^{ps} < 0$, as the flow of probability into state $i$ must be balanced by the flow of probability out of that state.
	
	As in our illustration, detailed balance implies further constraints. Summing over, and then integrating with time, the instantaneous currents along the edges of any cycle $c = \{i_1, \ldots, i_M\}$ we get (compare with equation \eqref{eq:kvl_example})
	\begin{equation}
	\label{eq:kvl}
	\sum_{j = 1}^M \Phi_{i_j i_{j+1}}^{ps} e^{B_{i_j i_{j+1}}} = 0 \quad,\quad \,i_{M+1} \equiv i_1.
	\end{equation}
	This implies that if one edge $(i_j, i_{j+1})$ of $c$ has $\Phi_{i_j, i_{j+1}}^{ps} > 0$ then there must exist at least one other edge $(i_k, i_{k+1})$ in $c$ with $\Phi_{i_k i_{k+1}}^{ps} < 0$. Thus, for any cycle, the non-zero $\Phi_{i_li_{l+1}}^{ps}$'s cannot all have the same sign. 	
	We now prove that \eqref{eq:kcl} and \eqref{eq:kvl} jointly imply $\Phi_{ij}^{ps} = 0$ for all edges. We establish this below by contradiction, assuming the existence of at least one edge $(m,n)$ with $\Phi_{mn}^{ps} > 0$.

\begin{figure}[tbp]
	\subfigure[\, Illustration of the construction of arrows. An arrow pointing along an edge, e.g. from $n$ to $m$, indicates a positive integrated probability current from $n$ to $m$, $\Phi_{mn}^{ps} > 0$.]{
	\label{fig:generala}
	\includegraphics[scale = .5, angle = 0]{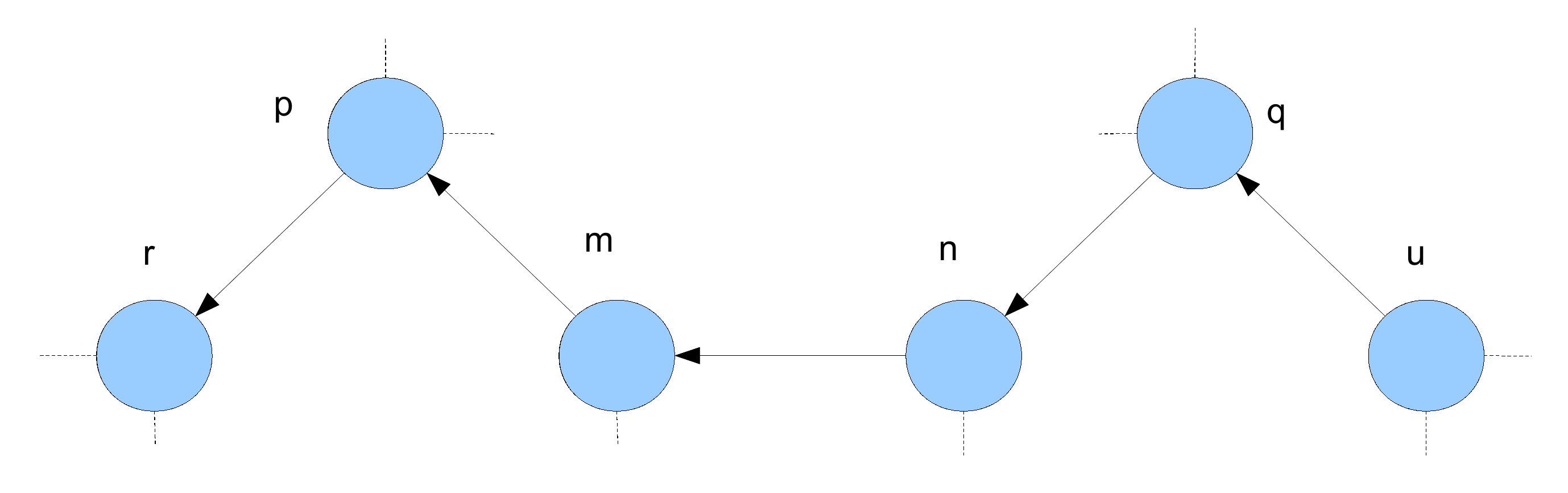}
	}
	\subfigure[\, One of the possible cycles, $\{m,p,q,n\}$, with all arrows pointing the same way.]{
	\label{fig:generalb}
	\includegraphics[scale = .5, angle = 0]{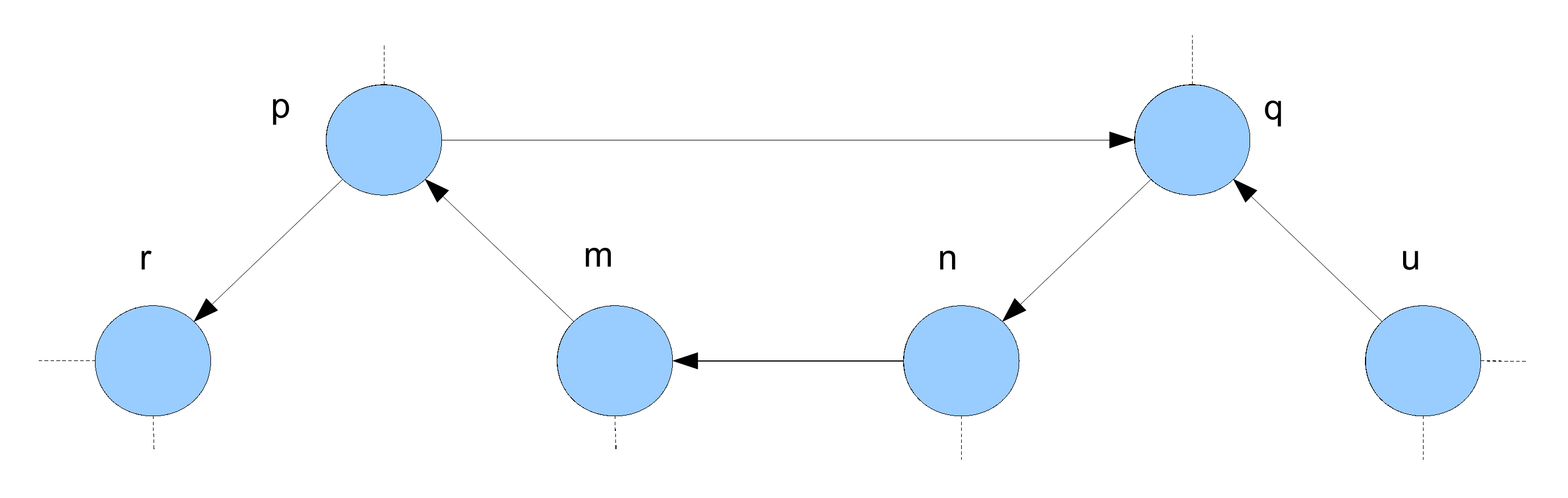}
	}
	\caption{\, Part of an $N$ state graph with arbitrary topology.}
	\label{fig:general}
	\end{figure}

	To formulate our argument, let us introduce the following convenient construction on $G$. Along every edge, say $(r,s)$, with non-zero $\Phi_{rs}^{ps}$, we draw an arrowhead indicating the positive direction of the integrated probability current, as shown in Fig. \ref{fig:generala}. By assumption, $G$ contains at least one arrow, pointing from $n$ to $m$. Equation \eqref{eq:kcl} then implies the existence of another edge $(p,m)$, such that $\Phi_{mp}^{ps} < 0$, or equivalently, $\Phi_{pm}^{ps} > 0$. Thus we must have another arrow pointing from $m$ to some $p  \neq n$. Similarly there must be another arrow from some vertex $q \neq m$ to $n$, to prevent the depletion of probability from state $n$. Refer to Fig. \ref{fig:generala} for illustration.
	
	Consider now the set $\mathcal{D}$ of all vertices that can be reached from $m$ by following the arrows. In Fig. \ref{fig:generala} $\mathcal{D} = \{p,r, \ldots\}$. Consider also set $\mathcal{S}$ of all vertices from which $n$ can be reached by following the arrows. In Fig. \ref{fig:generalb} $\mathcal{S} = \{q,u, \ldots\}$. These two sets must have at least one element in common, otherwise there will be a constant drainage of probability from $\mathcal{S}$ to $\mathcal{D}$ which is inconsistent with a periodic steady state. Let $v$ denote this common element. 
	
	The existence of a common element has an interesting consequence. Starting from state $m$, we can reach state $v$ by following the arrows (since $v \in \mathcal{D}$), and from there we can reach state $n$ by continuing to follow arrows (since $v \in \mathcal{S}$). Since an arrow points from $n$ to $m$, we conclude that there exists a cycle $\{m,..,v,..,n\}$ consisting of edges with arrows all pointing in the same direction $\{m \rightarrow .. \rightarrow v \rightarrow .. \rightarrow n \rightarrow m \}$. By construction, the $\Phi_{i_j i_{j+1}}^{ps}$'s along this cycle are all positive. One such cycle $\{m,p,q,n\}$ is shown in Fig. \ref{fig:generalb}.
	
	However, this contradicts equation \eqref{eq:kvl}. We conclude that the existence of a non-zero $\Phi_{mn}^{ps}$ is inconsistent with our starting assumptions, and this completes our proof.  	
	
	\section{Conclusion}
	\label{sec:conclusion}
	
	Recent interest and experimental progress in the synthesis of artificial molecular machines (see e.g.\ Refs.~\cite{Pei2006, Lund2010, Simmel2009,Leigh2003,Gu2010}) have stimulated basic theoretical work on the control of stochastic systems by the variation of external parameters~\cite{Astumian2007,Sinitsyn2007b,Rahav2008,Chernyak2008,Chernyak2009,Sinitsyn2009,Horowitz2009,Maes2010}.
	Among the results that have been obtained is the no-pumping theorem stated in Section~\ref{sec:setup}, which specifies minimal requirements for generating directed motion in periodically driven stochastic pumps.
	In this paper we have presented a simple proof of this theorem, based on the idea that if a non-zero integrated current is generated along some edge of the graph, then this edge must be part of a closed loop along which probability is conveyed in one direction: all the $\Phi_{i_j i_{j+1}}^{ps}$'s along the cycle have the same sign.
	This in turn is inconsistent with the assumption of detailed balance with fixed energy barriers (which gives equation \eqref{eq:kvl}).
	
	Very recently, Ren {\it et al}~\cite{Ren2011} have shown that when the network itself is a single cycle, the no-pumping theorem follows from a duality between state and barrier energies.
	Such a duality was previously noted by Jack and Sollich~\cite{Jack2009} for infinite one-dimensional lattice models.
	It would be interesting to see whether this duality can be extended to more complicated network topologies.
		
	\acknowledgments
	We gratefully acknowledge useful discussions with Debasis Mandal, Suriyanarayanan Vaikuntanathan and Jordan Horowitz, and financial support from the National Science Foundation (USA) under grants ECCS-0925365 and DMR-0906601, and the University of Maryland, College Park.

	\bibliography{dibref}
	\end{document}